\documentclass[letterpaper,english]{revtex4}
\usepackage{ae}
\usepackage{aecompl}
\usepackage[T1]{fontenc}
\usepackage[latin1]{inputenc}
\usepackage{subfigure}
\usepackage{graphicx}
\usepackage{amssymb}

\makeatletter

\usepackage{subfigure}
\usepackage{graphicx}

\usepackage{babel}
\makeatother
\begin{document}

\title{Multimode Interference: Identifying Channels and Ridges in Quantum
Probability Distributions}

\author{Ross C. O'Connell}

\email{rcoconne@umich.edu}

\affiliation{Department of Physics\\
University of Michigan\\
Ann Arbor, MI 48104}

\author{Will Loinaz}

\email{waloinaz@amherst.edu}

\affiliation{Department of Physics \\
 Amherst College \\
 Amherst, MA 01002}

\begin{abstract}
The multimode interference technique is a simple way to study the
interference patterns found in many quantum probability distributions.
We demonstrate that this analysis not only explains the existence
of so-called {}``quantum carpets,'' but can explain the spatial
distribution of channels and ridges in the carpets. With an understanding
of the factors that govern these channels and ridges we have a limited
ability to produce a particular pattern of channels and ridges by
carefully choosing the weighting coefficients $c_{n}$. We also use
these results to demonstrate why fractional revivals of initial wavepackets
are themselves composed of many smaller packets. 
\end{abstract}
\maketitle

\section{Introduction}

\newcommand{\Erf}[1]{\textrm{Erf}\left(#1\right)}

\newcommand{\re}[1]{\textrm{Re}\left\{ #1\right\} }

\newcommand{\im}[1]{\textrm{Im}\left\{ #1\right\} }

\newcommand{\row}[2]{\begin{}{array}[t]{c
 \begin{}{array}[t]{r}
 \left(#1\right.\\
 \left.#2\right).\end{}{array}
\end{}{array}
}}

\newcommand{\rowc}[2]{\begin{}{array}[t]{r
 \left(#1\right.\\
 \left.#2\right),\end{}{array}
}}

\newcommand{\Z}{\in \mathbb{Z}}

\newcommand{\p}[1]{\left(#1\right)}

The existence of multiple time scales in many simple, one-dimensional
quantum systems is both an important characteristic that distinguishes
them from their classical counterparts and the source of many interesting
phenomena in the time-evolution of those systems. Let us begin with
a wavefunction of the form \begin{equation}
\Psi(x,t)=\sum_{n=1}^{\infty}c_{n}\psi_{n}(x)e^{-iE_{n}t},\end{equation}
 where $\psi_{n}$ is the $n$th eigenfunction, $E_{n}$ the energy
of the $n$th eigenfunction, and the $c_{n}$ are weighting coefficients.
The most straightforward way to identify the time scales is to treat
$E_{n}$ as a function of $n$ and perform a Taylor expansion of the
energy around some quantum number $\bar{n}$, the number around which
the weighting coefficients are centered, \begin{equation}
E_{n}\approx E_{\bar{n}}+E_{\bar{n}}^{\prime}\left(n-\bar{n}\right)+\frac{1}{2!}E_{\bar{n}}^{\prime\prime}\left(n-\bar{n}\right)^{2}+...\end{equation}
 We identify the time scales as \begin{equation}
\frac{2\pi}{T_{j}}=\frac{E_{\bar{n}}^{(j)}}{\hbar j!}.\end{equation}
 The $T_{j}$ almost always obey the hierarchy $T_{1}\ll T_{2}\ll T_{3}...$,
as is demonstrated in Appendix \ref{sec:timescales}. $T_{1}$ corresponds
to the classical period of a particle of energy $E_{\bar{n}}$, so
we call it $T_{cl}$. At times near $T_{2}$ the contribution from
the $E_{\bar{n}}^{\prime\prime}\left(n-\bar{n}\right)$ terms to the
phase of each eigenfunction will be negligible. When the aforementioned
hierarchy holds, the contribution from the $T_{j}$ terms for $j>3$
will also be negligible and the classical behavior governed by $T_{cl}$
will dominate, approximately reproducing the $t\approx0$ behavior
of the particle. The return of the wavefunction to its $t\approx0$
value is called a quantum revival, and we call $T_{2}$ the revival
time, $T_{R}$. The literature on the subject of revivals is well-developed,
including several general articles \cite{rev:Aronstein2,rev:Bluhm1,rev:Rozmej1},
treatments of fractional revivals \cite{rev:Averbukh1,rev:Aronstein1,rev:Jie1,rev:Loinaz1,rev:Naqvi1},
and articles on possible applications of this phenomenon \cite{rev:Chen1,rev:Knopse1,misc:Averbukh1}.
Revivals are also analogous to the Talbot effect in classical electrodynamics,
as is discussed in \cite{opt:Berry1,opt:Berry2,opt:Dubra1,opt:Lock1}.

We wish to explain the behavior of the wavefunction at times comparable
to, but not equal to, $T_{R}$. At these times, in problems with anharmonic
spectrum and for wavefunctions $\Psi(x,t)$ in which the distribution
of $c_{n}$ has central value $\bar{n}$ and characteristic width
$\Delta n$ which obey the hierarchy $1\ll\Delta n\ll\bar{n}$ (a
semiclassical case), the spacetime diagram of the probability distribution
exhibits a complex pattern of channels and ridges -- what has come
to be called a quantum carpet. The problem has been approached with
several techniques \cite{car:Friesch1,car:Grossman1,car:Hall1,car:Marzoli1,car:Marzoli2},
and finds application in study of Bose-Einstein condensates \cite{bec:Choi1,bec:Schleich1,bec:Wright1}.
In this paper, we develop the technique of multimode interference
\cite{trace:Kaplan1,trace:Kaplan2}. In section \ref{sec:MMIandChVs}
we briefly review the multimode interference technique and introduce
characteristic velocities. In section \ref{sec:CharVs} we obtain,
in the context of the WKB approximation, the characteristic velocities
of channels and ridges in the spacetime diagram of the probability
distribution, and in section \ref{sec:Groups-of-Velocities} we examine
the way in which the degeneracy of these velocities shapes the quantum
carpet. These ideas are illustrated with the example of the infinite
square well potential in section \ref{sec:ISW} and are used to elucidate
the structure of fractional revivals in section \ref{sec:psi-cl}.
Section \ref{sec:Summary} offers a summary of lessons learned.

\section{\label{sec:MMIandChVs}Multimode Interference and Characteristic
Velocities}

A multimode term is a straightforward adaptation of a density matrix
element,\begin{eqnarray}
\mu_{nm}(x,t) & = & d_{nm}\psi_{n}(x)\psi_{m}^{*}(x)e^{-it\left(E_{n}-E_{m}\right)},\label{eq:intermode-def}\\
d_{nm} & = & c_{n}c_{m}^{*}=d_{mn}^{*},\end{eqnarray}
 which allows us to write a probability density as \begin{equation}
\left\Vert \Psi(x,t)\right\Vert ^{2}=\sum_{n,m=1}^{\infty}\mu_{nm}(x,t).\end{equation}
 While the multimode terms lack some of the convenient properties
of the density matrix they can simplify the study of some dynamic
phenomena, since they allow us to write the probability density as
a simple sum. Of course, in order to do much with them we will have
to make an assumption about the eigenfunctions.

We will use the WKB approximation \cite{misc:Griffiths1} to obtain
a usable form of the eigenfunctions, $\psi_{n}$. While this limits
the range of applicability of our findings, quantum probability distributions
only acquire the features that earn the name {}``quantum carpet''
in the semiclassical case %
\footnote{The term {}``semiclassical'' is a bit ambiguous. Although a distribution
of weighting coefficients $c_{n}$ that satisfies $1\ll\Delta n\ll\bar{n}$
seems to be most important in producing semiclassical behavior we
are here willing to relax this to the WKB assumption that the potential
changes much more slowly than the relevant eigenfunctions oscillate.
Note that $\Delta n\gg1$ implies $\Delta p\gg1,$ which in turn suggests
(via the uncertainty principle) that $\Delta x\ll1,$ and thus that
the WKB approximation is appropriate.%
}. The approximate eigenfunctions are \begin{equation}
\psi_{n}(x)\approx\frac{C_{n}^{(-)}}{\sqrt{p_{n}(x)}}\exp\left(i\int^{x}p_{n}\left(x^{\prime}\right)dx^{\prime}\right)+\frac{C_{n}^{(+)}}{\sqrt{p_{n}(x)}}\exp\left(-i\int^{x}p_{n}\left(x^{\prime}\right)dx^{\prime}\right),\label{eq:psi-wkb}\end{equation}
 where the $C_{n}^{(\pm)}$ are complex constants of integration,
and \begin{equation}
p_{n}(x)=\sqrt{E_{n}-V(x)}.\end{equation}
 The approximate multimode terms $\mu_{nm}$ will be weighted sums
of four intermode terms, \begin{equation}
\mu_{nm}\approx d_{nm}\left(\iota_{nm}^{(++)}+\iota_{nm}^{(+-)}+\iota_{nm}^{(-+)}+\iota_{nm}^{(--)}\right),\end{equation}
 where\begin{equation}
\iota_{nm}^{(\pm_{1}\pm_{2})}(x,t)=d_{nm}\frac{C_{n}^{(\pm_{1})}C_{m}^{(\pm_{2})}}{\sqrt{p_{n}(x)\, p_{m}(x)}}\exp-i\left(\int^{x}\left(\pm_{1}p_{n}\left(x^{\prime}\right)\pm_{2}p_{m}\left(x^{\prime}\right)\right)dx^{\prime}+\left(E_{n}-E_{m}\right)t\right).\label{eq:wkb-phase}\end{equation}
 We are interested in features of the probability density that are
time-dependent and typically much more {}``narrow'' than the potential,
so we focus our attention on the phase factor instead of the leading
factors of $p_{n}^{-1/2}$. We can characterize the curves of constant
phase by differentiating the argument of the exponential and looking
for its roots, \begin{eqnarray}
\frac{d}{dt}\left(\int^{x\left(t\right)}\left(\pm_{1}p_{n}\left(x^{\prime}\right)\pm_{2}p_{m}\left(x^{\prime}\right)\right)dx^{\prime}+\left(E_{n}-E_{m}\right)t\right) & = & 0,\\
\frac{dx}{dt} & = & -\frac{E_{n}-E_{m}}{\pm_{1}p_{n}\left(x\right)\pm_{2}p_{m}\left(x\right)},\\
v_{nm}^{(\pm_{1}\pm_{2})}=\frac{\Delta\omega}{\Delta k} & = & -\frac{E_{n}-E_{m}}{\pm_{1}\sqrt{E_{n}-V(x)}\pm_{2}\sqrt{E_{m}-V(x)}},\label{eq:vel-def}\end{eqnarray}
 we find that each pair of quantum numbers, $\left(n,m\right)$, gives
rise to \emph{four} velocities $v_{nm}^{(\pm_{1}\pm_{2})}$. We can
pick any one of these velocities, integrate it from an arbitrary $x_{0}$,
and the resulting trajectory is a line of constant phase. Channels
and ridges in the probability distribution follow these lines, as
we will demonstrate in section \ref{sec:Groups-of-Velocities}.

\section{\label{sec:CharVs}Characterization of the Velocities}

We have written eq. \ref{eq:vel-def} as $\Delta\omega/\Delta k$
anticipating that we will label some of the velocities as {}``group
velocities.'' Defining $\omega_{n}=E_{n}$ and $k_{n}(x)=p_{n}(x)$,
we can write \begin{equation}
\omega_{n}(k_{n},x)=k_{n}^{2}-V(x).\end{equation}
 We now make the additional assumption that the weighting coefficients
are well-centered on some number $\bar{n}$ with some spread $\Delta n$,
and that these satisfy the hierarchy $1\ll\Delta n\ll\bar{n}$. This
allows us to define a group velocity for our packet, \begin{equation}
v_{gr}=\left.\frac{d\omega_{n}}{dk_{n}}\right|_{n=\bar{n}}=2k_{\bar{n}},\end{equation}
 which is the classical velocity of a particle in the potential $V$
with energy $E_{\bar{n}}$.

Writing $E_{n}=E_{\bar{n}}+e_{n}$ and $E_{m}=E_{\bar{n}}+e_{m}$,
where $e_{n},e_{m}\ll E_{\bar{n}}-V(x)$, we can simplify our expression
for the velocities in eq. \ref{eq:vel-def},\begin{eqnarray}
v_{nm}^{(\pm_{1}\pm_{2})} & = & -\frac{\left(E_{\bar{n}}+e_{n}\right)-\left(E_{\bar{n}}+e_{m}\right)}{\pm_{1}\sqrt{E_{\bar{n}}+e_{n}-V(x)}\pm_{2}\sqrt{E_{\bar{n}}+e_{m}-V(x)}}\nonumber \\
 & \approx & -\frac{e_{n}-e_{m}}{\sqrt{E_{\bar{n}}-V(x)}\left(\pm_{1}\left(1+\frac{e_{n}}{2\left(E_{\bar{n}}-V(x)\right)}\right)\pm_{2}\left(1+\frac{e_{m}}{2\left(E_{\bar{n}}-V(x)\right)}\right)\right)}\nonumber \\
 & \approx & -2\sqrt{E_{\bar{n}}-V(x)}\frac{e_{n}-e_{m}}{2\left(E_{\bar{n}}-V(x)\right)\left(\pm_{1}1\pm_{2}1\right)+\left(\pm_{1}e_{n}\pm_{2}e_{m}\right)}\nonumber \\
 & \approx & \left\{ \begin{array}{l}
\pm2\sqrt{E_{\bar{n}}-V(x)}=\pm_{1}2k_{\bar{n}}(x),\,\left(+_{1}-_{2},\,-_{1}+_{2}\right)\\
\pm\frac{e_{n}-e_{m}}{2\sqrt{E_{\bar{n}}-V(x)}}\approx\pm\frac{\omega_{n}-\omega_{m}}{2k_{\bar{n}}(x)},\,\left(+_{1}+_{2},\,-_{1}-_{2}\right).\end{array}\right.\label{eq:2-vel}\end{eqnarray}
 As long as both $e_{n},e_{m}\ll E_{\bar{n}}-V(x)$, as is required
in the semiclassical approximation, then half of the velocities contributed
by a particular $\left(n,m\right)$ are comparable to the group velocity,
$v_{gr}$, and half are smaller. It is important to note (in preparation
for the next section) that eq. \ref{eq:2-vel} is approximate -- it
is \emph{not} the case that half of the velocity terms are exactly
degenerate.

\section{\label{sec:Groups-of-Velocities}Groups of Velocities and Degeneracy}

The result in eq. \ref{eq:2-vel} demonstrates that in any problem
there exists a range of velocities. Though that result was approximate,
we will now show that some intermode terms have \emph{exactly} the
same maximum velocities, and that because of this we can treat those
intermode terms as essentially moving together. This is a sort of
degeneracy -- the sort of degeneracy that produces quantum carpets.

If we examine the velocities (eq. \ref{eq:vel-def}) at a point where
$V(x)=0$, we find that\begin{equation}
v_{nm}^{(\pm_{1}\pm_{2})}=-\frac{E_{n}-E_{m}}{\pm_{1}\sqrt{E_{n}}\pm_{2}\sqrt{E_{m}}}.\end{equation}
 We can factor the numerator,\begin{equation}
v_{nm}^{(\pm_{1}\pm_{2})}=-\frac{\left(\sqrt{E_{n}}+\sqrt{E_{m}}\right)\left(\sqrt{E_{n}}-\sqrt{E_{m}}\right)}{\pm_{1}\sqrt{E_{n}}\pm_{2}\sqrt{E_{m}}},\end{equation}
 and arrive at an important condition,\begin{equation}
v_{nm}^{(\pm_{1}\pm_{2})}=\mp_{1}\sqrt{E_{n}}\pm_{2}\sqrt{E_{m}}.\label{eq:degen-cond}\end{equation}
 Terms that have the same $v_{nm}^{(\pm_{1}\pm_{2})}$ at $V(x)=0$
change phase with the same period -- these are our degenerate terms.

We define a group of velocities as\begin{equation}
\beta_{v}(x,t)=\sum_{\left|v_{nm}^{(\pm_{1}\pm_{2})}\right|=v}\iota_{nm}^{(\pm_{1}\pm_{2})}(x,t).\end{equation}
 Because the first sign in the definition of the velocities (eq. \ref{eq:degen-cond})
is flipped relative to the signs in the definition of an intermode
term (eq. \ref{eq:wkb-phase}), each $\beta_{v}$will look roughly
like a weighted sum of harmonic functions. If there are many terms
in the sum, they may interfere to produce a peak or a valley. As the
wavefunction evolves in time, these peaks and valleys will trace out
the canals and ridges that form the quantum carpet.

\begin{figure}
\includegraphics{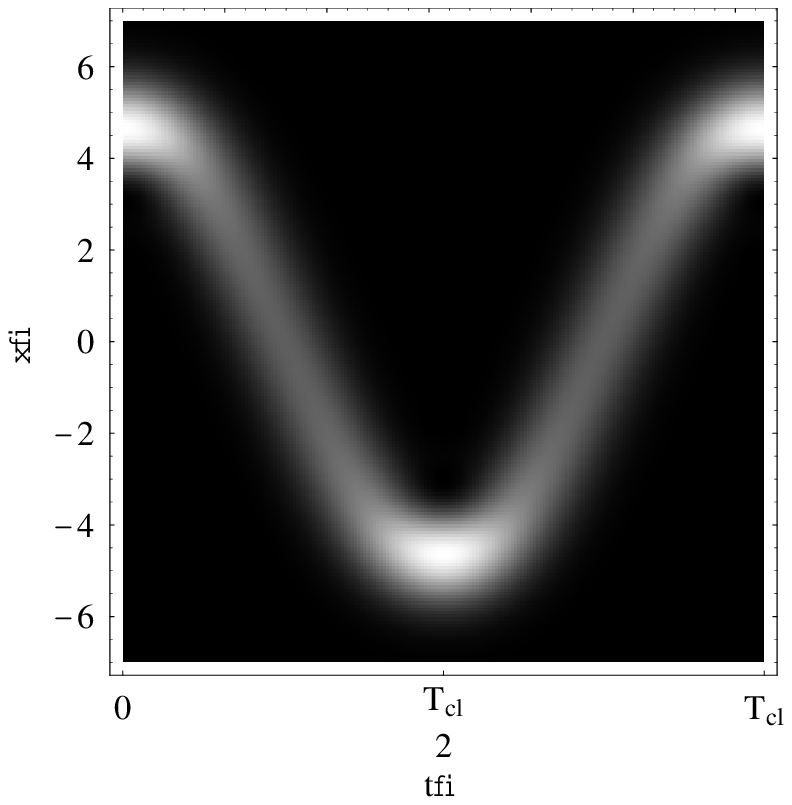}\hfill{}\includegraphics{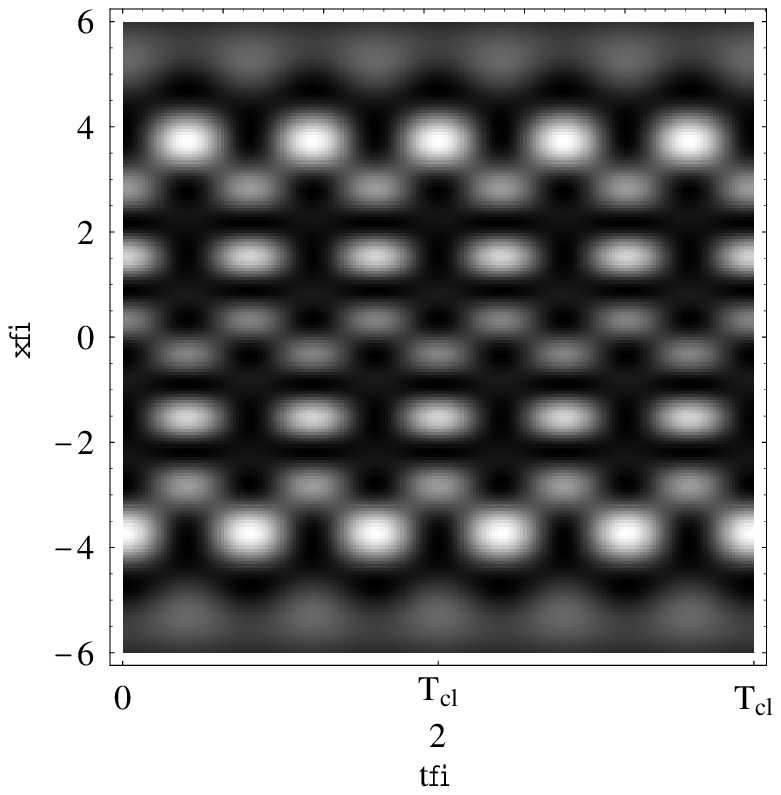}

\caption{\label{fig: sho-sqr}Two plots of the simple harmonic oscillator.
On the left, the distribution of coefficients is Gaussian, with $\bar{n}=6$
and $\sigma_{n}=2$, while the plot on the right is an even weighting
of the perfect squares between 1 and 81.}
\end{figure}

This immediately demonstrates the role of quadratic spectra in producing
quantum carpets. If the spectrum depends on the quantum number $n$
squared, then there will be many degenerate velocities, while if the
spectrum is linear in the quantum number, only traces which involve
two perfect squares may be degenerate. This suggests that if we begin
with a system like the simple harmonic oscillator, with a spectrum
linear in the quantum number, and set to zero all weighting coefficients
that are not perfect squares (selecting only states 1,4,9,16, etc.),
we can produce a carpet because we have effectively {}``quadratized''
the spectrum %
\footnote{{}``Effectively quadratize the spectrum'' is shorthand for choosing
weighting coefficients such that only terms that have quantum numbers
that are perfect squares are contained in the wavefunction. When we
do this, we could rewrite the spectrum, eigenfunctions, etc., as if
they were governed by a new variable, $m^{2}=n$, as if they had quadratic
spectra. Of course, we're not changing the spectrum itself, we're
changing the wavefunction.%
}. For an example of this, see Figure \ref{fig: sho-sqr}.%

\section{\label{sec:ISW}The Infinite Square Well}

\begin{figure}
\includegraphics{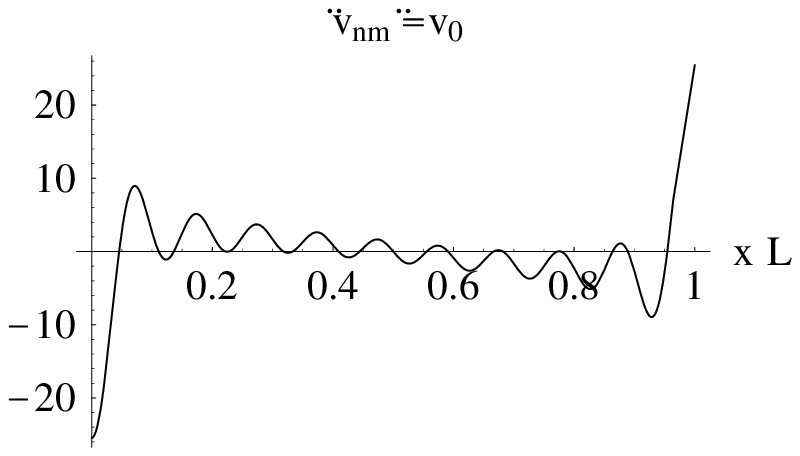}\hfill{}\includegraphics{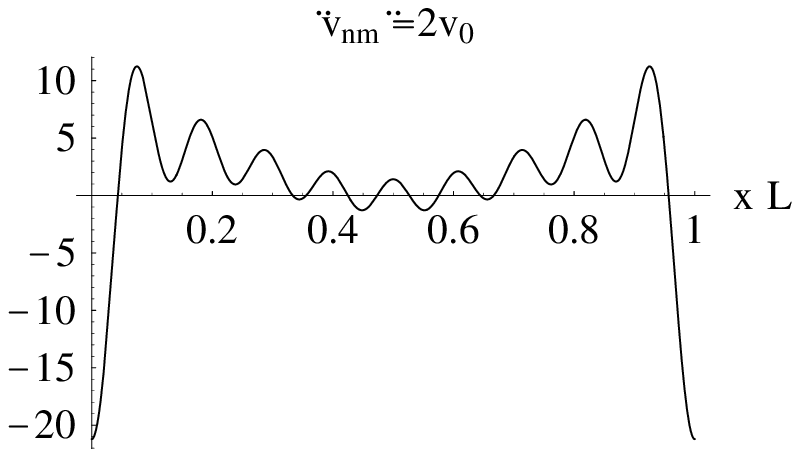}

\caption{\label{fig:trace-v}The $\left|v_{nm}\right|=v_{0}$ (left) and $\left|v_{nm}\right|=2v_{0}$
(right) velocity bundles for a uniform distribution of weighting coefficients
between $n=1$ and $n=10$ in the infinite square well.}
\end{figure}

We now turn our attention to a particular example in order to illustrate
the utility of these results. We write explicitly all of the relevant
physical constants, so as to facilitate comparison with classical
results. For a particle of mass $M$ in a well of width $L$ we have
the wavefunction,

\begin{equation}
\Psi\left(x,t\right)=\sqrt{\frac{2}{L}}\sum_{n=1}^{\infty}c_{n}\frac{i}{2}\left(\exp\left(i\frac{n\pi}{L}x\right)-\exp\left(-i\frac{n\pi}{L}x\right)\right)\exp\left(-i\frac{\hbar\pi^{2}n^{2}}{2ML^{2}}t\right).\end{equation}
 We can immediately identify the velocities in question as \begin{equation}
v_{nm}^{\pm_{1}\pm_{2}}=\frac{\hbar\pi}{2ML}\left(\mp_{1}n\pm_{2}m\right)=v_{0}\left(\mp_{1}n\pm_{2}m\right).\end{equation}
 We can also find that the wavenumbers $k_{nm}$ will be \begin{equation}
k_{nm}^{\pm_{1}\pm_{2}}=\frac{\pi}{L}\left(\pm_{1}n\pm_{2}m\right),\end{equation}
 so that in any particular velocity bundle there will be a variety
of wavenumbers. We provide an example of the interference that this
produces in figure \ref{fig:trace-v}.%

\begin{figure}
\includegraphics[%
  width=0.25\paperwidth]{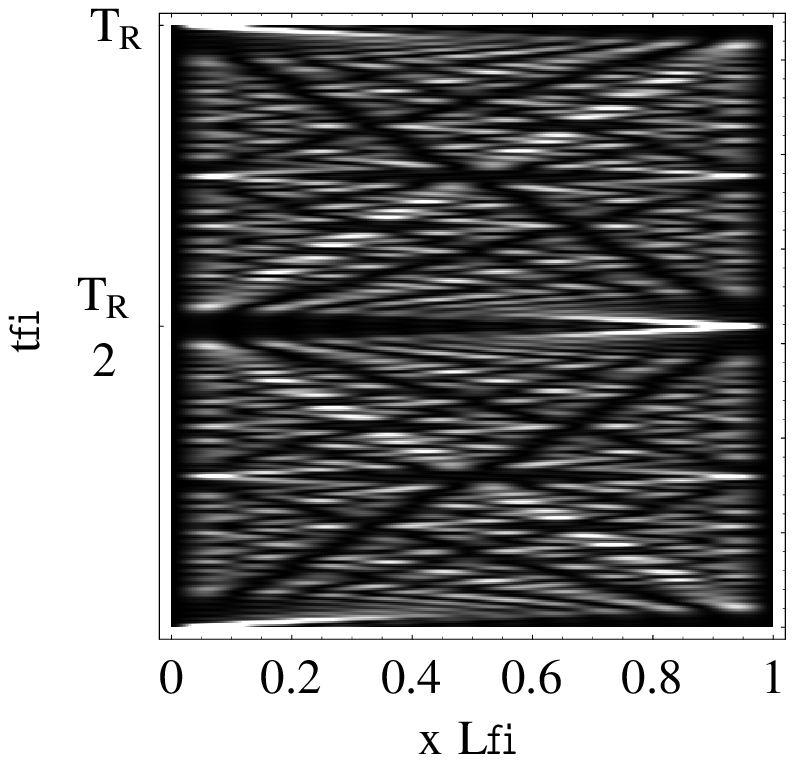}\hfill{}\includegraphics[%
  width=0.25\paperwidth]{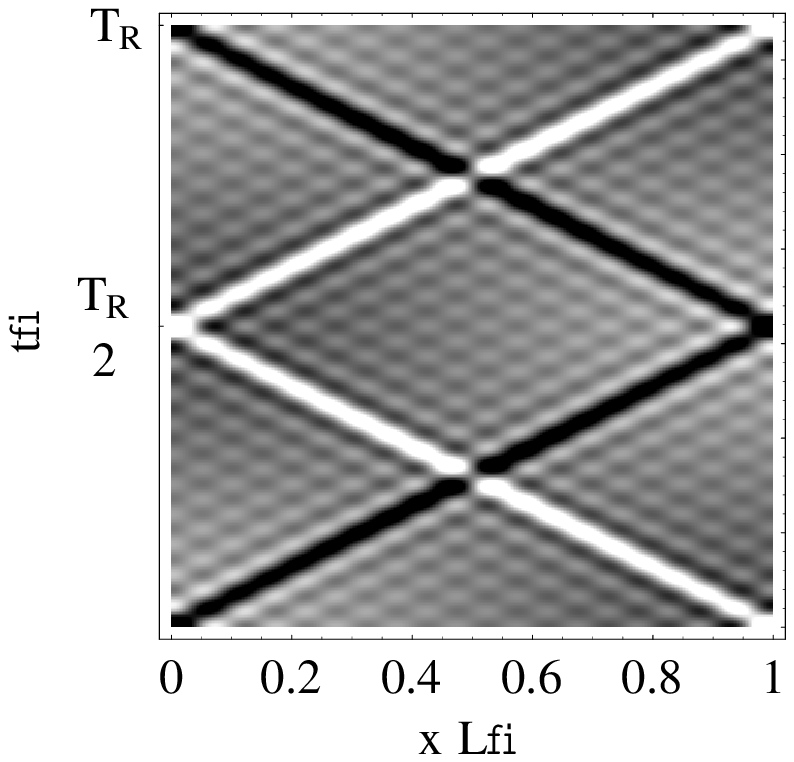}\hfill{}\includegraphics[%
  width=0.25\paperwidth]{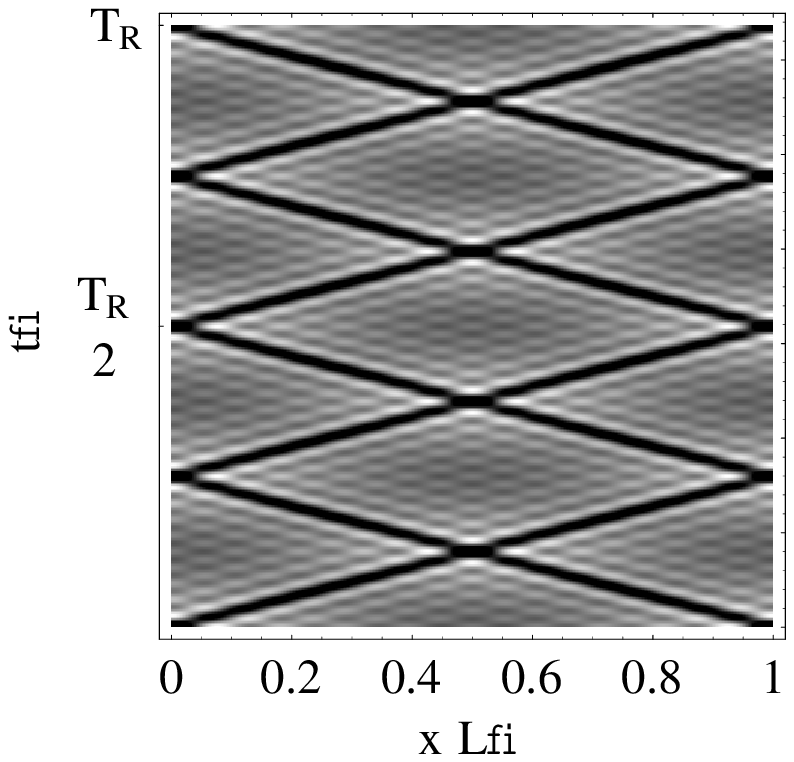}

\caption{\label{fig:carpet-0}At left, the full carpet, coefficients evenly
weighted between $n=1$ and $n=10$. In the middle, the eighteen intermode
terms that compose $\beta_{v_{0}}$. Note that they do, indeed, have
a period $T=T_{R}$. At right the 34 intermode terms that compose
$\beta_{2v_{0}}$. Note that they have a period $T=T_{R}/2$.}
\end{figure}

The most degenerate velocity will typically be $v_{0}$, and in time
$T_{R}=2L/v_{0}$ a trajectory with this velocity will cover a distance
of $2L$, so the most prominent traces should have the same period
as the revival time. Better still, we can ask what velocity we need
in order to have a period equal to the classical period, $T_{cl}$.
The condition is \begin{eqnarray}
v_{nm}T_{cl}=v_{nm}\frac{L}{\bar{n}v_{0}} & = & 2L\nonumber \\
v_{nm} & = & 2\bar{n}v_{0},\end{eqnarray}
 which should be satisfied by very few $\p{n,m}$ pairs. An example
of how well our separation of the wavefunction works is shown in Figure
\ref{fig:carpet-0}.%

In the case of the square well we can find a closed form for $\beta_{v}(x,0)$.
From the definition and the form of $v_{nm}^{(\pm_{1}\pm_{2})}$ it
can be shown that\begin{eqnarray*}
\beta_{v}(x,t) & = & \left(-\sum_{n=1}^{\infty}\frac{c_{n}c_{n+v}^{*}}{4}\left(\iota_{n(n+v)}^{(++)}(x,t)+\iota_{n(n+v)}^{(--)}(x,t)\right)-\sum_{m=1}^{\infty}\frac{c_{m+v}c_{m}^{*}}{4}\left(\iota_{(m+v)m}^{(++)}(x,t)+\iota_{(m+v)m}^{(--)}(x,t)\right)\right.\\
 &  & \hfill\left.+\sum_{n=1}^{v-1}\frac{c_{n}c_{v-n}^{*}}{4}\left(\iota_{n(v-n)}^{(+-)}(x,t)+\iota_{n(v-n)}^{(-+)}(x,t)\right)\right)\end{eqnarray*}
 and recalling that the $\iota_{nm}^{(\pm_{1}\pm_{2})}$ are just
exponentials, we can write $\beta_{v}(x,0)$ as \begin{equation}
\beta_{v}(x,t)=\frac{2}{L}\sum_{n=1}^{\infty}\textrm{Re}\left\{ c_{n}c_{n+v}^{*}\right\} \cos\left((2n+v)\frac{\pi x}{L}\right)+\frac{1}{L}\cos\left(v\frac{\pi x}{L}\right)\sum_{n=1}^{v-1}c_{n}c_{v-n}^{*}.\end{equation}
 The first term contributes peaks and valleys, while the second sets
a threshold height above which those peaks and valleys must rise in
order to be visible. The first term suggests that we can, by choosing
appropriate $c_{n}$, control the distribution of peaks and valleys,
though we cannot generate an arbitrary function %
\footnote{Note here that a choice of ${1,1,-1,-1,1,1,-1,-1}$ for the $c_{n}$,
with $c_{n}=0$ for $n>8$, yields a $\beta_{v_{0}}(x,0)$ with a
peak and a valley at the center of the well, and $\beta_{v_{0}}=0$
at the walls of the well. This contradicts the assertion about the
distribution of peaks and valleys contained in \cite{trace:Kaplan2}.%
}.

\section{\label{sec:psi-cl}Multimode Traces and Quantum Revivals}

In problems where the weighting coefficients $c_{n}$ are well-localized
around some central quantum number $\bar{n}$, so that the spectrum
can be well approximated by \begin{equation}
E_{n}\approx E_{\bar{n}}+E_{\bar{n}}^{\prime}\left(n-\bar{n}\right)+\frac{1}{2}E_{\bar{n}}^{\prime\prime}\left(n-\bar{n}\right)=E_{\bar{n}}+\frac{2\pi}{T_{cl}}k+\frac{2\pi}{T_{R}}k^{2},\end{equation}
 we can make an interesting approximation to the wavefunction at rational
fractions of $T_{R}$. We do not present the details of the derivation
here (see \cite{rev:Averbukh1} or \cite{O'Connell2002}), but quote
the result that \begin{equation}
\Psi\left(x,t\approx\frac{p}{q}T_{R}\right)\approx\sum_{s=0}^{l-1}a_{s}\Psi_{cl}\left(x,t+\frac{s}{l}T_{cl}\right),\end{equation}
 where $l=q/2$ when $q$ contains more than one power of two, $l=q$
otherwise, the $a_{s}$ are complex weighting coefficients (for their
values, see \cite{rev:Averbukh1} or \cite{O'Connell2002}), and we
have defined \begin{equation}
\Psi_{cl}\left(x,t\right)=\sum_{k=-\infty}^{\infty}c_{k+\bar{n}}\psi_{k+\bar{n}}(x)\exp\left(-2\pi ik\frac{t}{T_{cl}}\right).\end{equation}
 This approximation states that the wavefunction can be written as
the sum of slices of a {}``classicized'' wavefunction, $\Psi_{cl}$.
It suggests that at rational fractions of $T_{R}$ (particularly fractions
with low $q$), a wavefunction that began as a well-localized packet
will consist of several packets. We will now demonstrate that if $\Psi(x,t=0)$
is well-localized, then $\Psi_{cl}$ will remain well-localized for
all $t$.

Observe from the derivation of Equation \ref{eq:vel-def} that the
$E_{n}-E_{m}$ term comes from the time-evolution exponential and
the square root terms come from the WKB approximation. If we want
to find a similar formula for $\Psi_{cl}$, we need only replace the
$E_{n}-E_{m}$ in the numerator with $\left(2\pi/T_{1}\right)\left(n-m\right)$.
Considering a point where $V(x)=0$, we find the following velocity
degeneracy condition:\begin{equation}
v_{nm}=\pm_{1}\frac{2\pi}{T_{cl}}\frac{n-m}{\sqrt{E_{n}}\pm_{2}\sqrt{E_{m}}}.\end{equation}
 Unlike Equation \ref{eq:degen-cond}, we have no hope of factoring
this in general. We can, however, see what happens in a few obvious
cases. If \begin{equation}
E_{n}=\alpha^{2}n,\end{equation}
 then we quickly find \begin{equation}
v_{nm}=\pm_{1}\frac{2\pi}{T_{cl}\alpha}\frac{n-m}{\sqrt{n}\pm_{2}\sqrt{m}},\end{equation}
 a degeneracy condition identical to the one we would have found for
the harmonic oscillator in the previous section. If, however, we try
a spectrum $E_{n}=\alpha^{2}n^{2},$ we find something far more interesting,\begin{equation}
v_{nm}=\pm_{1}\frac{2\pi}{T_{cl}\alpha}\frac{n-m}{n\pm_{2}m}=\left\{ \begin{array}{l}
\pm_{1}\frac{2\pi}{T_{cl}\alpha},\,\left(-_{2}\right),\\
\pm_{1}\frac{2\pi}{T_{cl}\alpha}\frac{n-m}{n+m},\,\left(+_{2}\right).\end{array}\right.\end{equation}
 \emph{Half} of all of the traces are degenerate! Looking at the other
half, we want to find two pairs of numbers, $(n,m)$ and $(p,q)$
that will be degenerate. The conditions are \begin{equation}
\frac{n-m}{n+m}=\frac{p-q}{p+q},\,\begin{array}{cc}
n+m\neq0, & p+q\neq0,\\
n\neq p, & m\neq p,\end{array}\end{equation}
 which then reduce to \begin{eqnarray}
\left(p+q\right)\left(n-m\right) & = & \left(n+m\right)\left(p-q\right),\nonumber \\
qn-mp & = & -qn+mp,\nonumber \\
\frac{n}{m} & = & \frac{p}{q}.\end{eqnarray}
 While this is not a difficult equation to satisfy with all of the
integers at our disposal, it is difficult to satisfy when our weighting
coefficients are all within some $\Delta n\ll\bar{n}$ of $\bar{n}\gg1$
. The solutions that will lie closest to $\left(n,m\right)$ are $\left(p,q\right)=1/2\left(n,m\right)$
and $\left(p,q\right)=2\left(n,m\right)$. If $n$ and $m$ are in
the vicinity of $\bar{n}$, as they must be in the semiclassical case,
that puts $p$ and $q$ near either $\bar{n}/2$ or $2\bar{n}$, both
of which would typically be {}``out of range'' of $\Delta n$. For
the case of a quadratic spectrum, then, we find that half of the velocities
are degenerate, and that the other half tend to be non-degenerate.
This nicely corresponds with our suspicion that $\Psi_{cl}$ wavepackets
evolve along classical paths.

\begin{figure}
\subfigure[The Morse Potential]{\includegraphics[%
  width=0.45\textwidth]{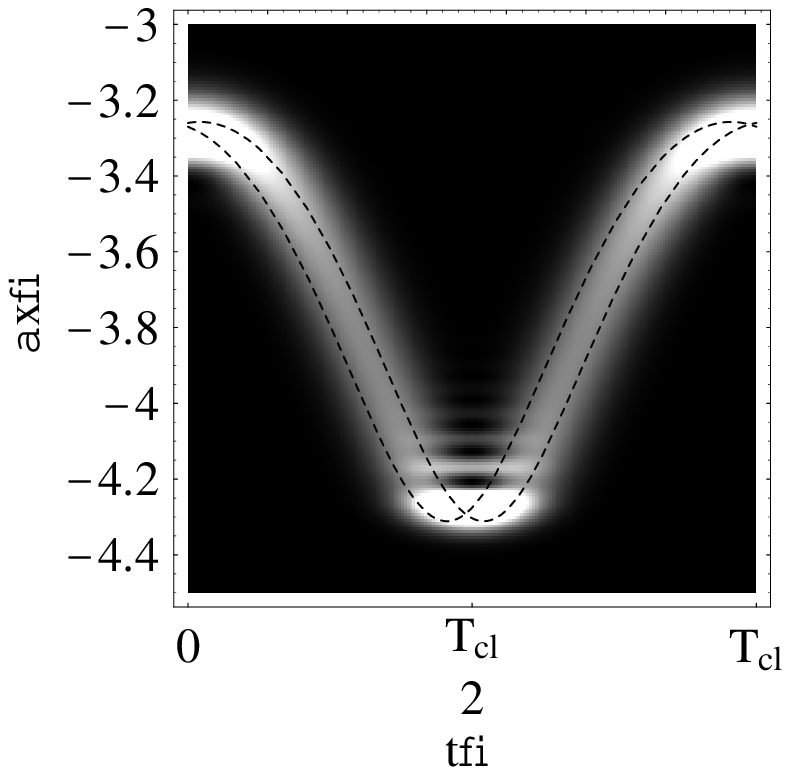}}\hfill{}\subfigure[The Rosen-Morse I Potential]{\includegraphics[%
  width=0.44\textwidth]{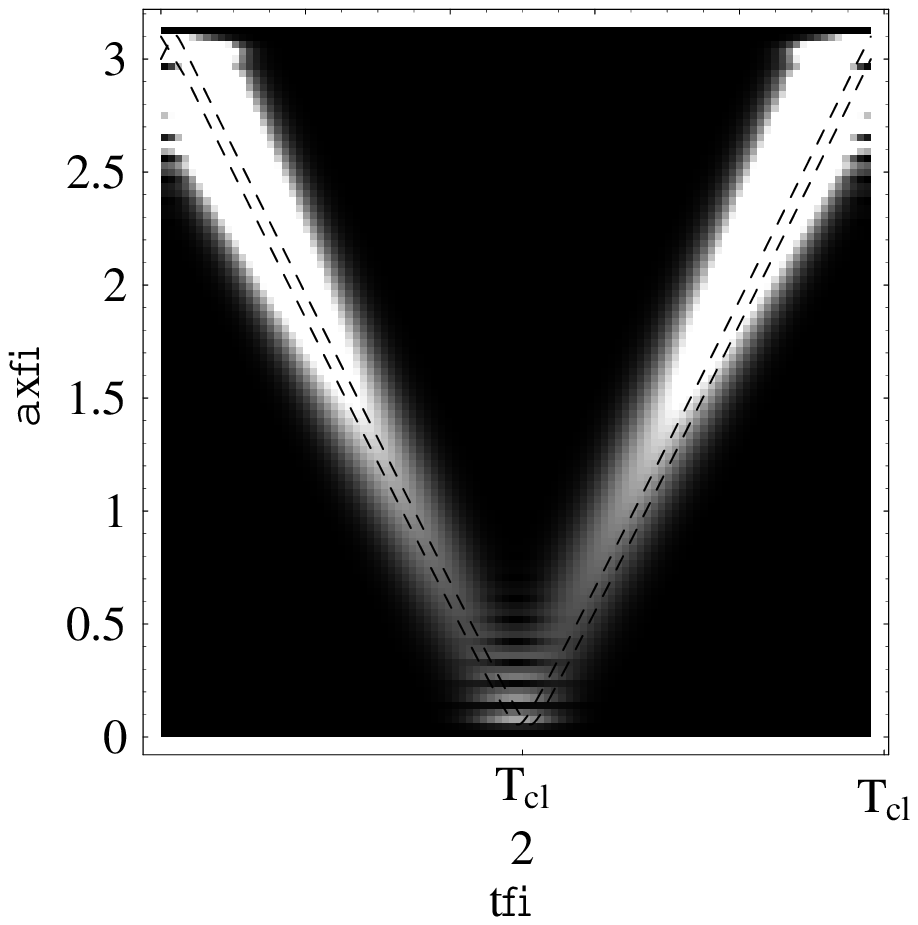}}

\caption{\label{fig: quad/non}Plots of $\Psi_{cl}$ in two potentials, one
with a quadratic spectrum and one with a non-quadratic spectrum. On
the left is the Morse Potential, with $E_{n}=A^{2}-\left(A-\alpha n\right)^{2}$,
and on the right the Rosen-Morse I potential, with $E_{n}=-A^{2}+\left(A+\alpha n\right)^{2}-\left(B/\left(A+\alpha n\right)\right)^{2}+\left(B/A\right)^{2}$.
In both cases $A,\, B,\,\textrm{and}\,\alpha$ are independent constants.
The black lines overlaying the maxima are classical paths for those
potentials with $E=E_{\bar{n}}$. These are exactly solvable potentials
taken from \cite{misc:Cooper1}.}
\end{figure}

The unfortunate aspect of this result for $\Psi_{cl}$ is that it
is so heavily dependent on the quadratic character of the spectrum.
This behavior seems to occur in potentials with non-quadratic spectra
as well, as demonstrated in Figure \ref{fig: quad/non}.%
However, in the semiclassical limit (that is, for very large $\bar{n}$),
their spectra can appear to be no more than quadratic \cite{misc:Nieto1}.
This leaves low-$\bar{n}$ cases to consider, and completing our understanding
of $\Psi_{cl}$ is a possible subject for future work.

\section{\label{sec:Summary}Summary}

We have seen how an analysis of the intermode terms in the probability
density of a semiclassical wavefunction can give us insight into the
degeneracy conditions which must be met in order to produce a carpet,
and into the carpet itself. When those degeneracy conditions are met,
we can examine the initial form of certain bundles of velocities and
from that find the channels and ridges of the quantum carpet. For
a problem with a spectrum linear in the quantum number it was almost
impossible to generate a carpet, though a perverse choice of weighting
coefficients could effectively quadratize the spectrum. Similar reasoning
suggested why $\Psi_{cl}$ should so often resemble a classically
oscillating packet, and thus why fractional revivals of a wavepacket
should themselves resemble the initial packet.

\appendix

\section{\label{sec:timescales}Demonstration that $T_{1}\ll T_{2}\ll T_{3}...$}

We begin by assuming that our spectrum can be written as a polynomial
in $n$,\begin{equation}
E_{n}=\sum_{m=1}^{M}a_{m}n^{m}.\end{equation}
 In this case, the $j$th derivative of $E_{n}$ is\begin{equation}
E_{n}^{(j)}=\sum_{m=j}^{M}a_{m}n^{m}\left(\frac{m!}{(m-j)!n^{j}}\right),\end{equation}
 and if $n>m-j$ the $j$th derivative will reduce the contribution
from the $n$th term. So long as $\bar{n}\gg M$, it will be true
that $E_{n}^{(1)}\gg E_{n}^{(2)}\gg E_{n}^{(3)}...$, since $\frac{m!}{\bar{n}!}<\frac{M!}{\bar{n}!}\ll1$.
Consequently, $T_{1}\ll T_{2}\ll T_{3}...$ holds. Our result is not
significantly changed by allowing negative powers of $n$. Consider\begin{equation}
E_{n}=\sum_{m=1}^{M}a_{m}n^{-m}.\end{equation}
 Its derivatives are \begin{equation}
E_{n}^{(j)}=\sum_{m=1}^{M}a_{m}n^{-m}(-1)^{j}\left(\frac{(m+j-1)!}{(m-1)!n^{j}}\right),\end{equation}
 so we still get that $T_{1}\ll T_{2}\ll T_{3}...$ so long as $\bar{n}\gg M$.
Transcendental spectra still present a problem, though we note that
the spectrum of \emph{any} one-dimensional potential cannot increase
faster than $n^{2}$ for large $n$, roughly because the walls of
the potential cannot be {}``harder'' than the infinite square well
\cite{misc:Nieto1}.

\bibliographystyle{apalike}
\bibliography{thesis-article}

\begin{thebibliography}{}

\bibitem[Aronstein, 2000]{rev:Aronstein2}
Aronstein, D.~L. (2000).
\newblock Analytical investigation of revival phenomena in the finite
  square-well potential.
\newblock {\em Physical Review A}, 62:022102.

\bibitem[Aronstein and Stroud, 1997]{rev:Aronstein1}
Aronstein, D.~L. and Stroud, C. R.~J. (1997).
\newblock Fractional wave-function revivals in the infinite square well.
\newblock {\em Physical Review A}, 55(6):4526--4537.

\bibitem[Averbukh and Perelman, 1989]{rev:Averbukh1}
Averbukh, I.~S. and Perelman, N. (1989).
\newblock Fractional revivals: Universality in the long-term evolution of
  quantum wave packets beyond the correspondence principle dynamics.
\newblock {\em Physics Letters A}, 139(9):449--453.

\bibitem[Averbukh and Perelman, 1991]{misc:Averbukh1}
Averbukh, I.~S. and Perelman, N. (1991).
\newblock The dynamics of wave packets of highly-excited states of atoms and
  molecules.
\newblock {\em Sov. Phys. Usp.}, 34(7):572--591.

\bibitem[Berry and Bodenschatz, 1999]{opt:Berry2}
Berry, M.~V. and Bodenschatz, E. (1999).
\newblock Caustics, multiply reconstructed by talbot interference.
\newblock {\em Journal of Modern Optics}, 46(2):349--365.

\bibitem[Berry and Klein, 1996]{opt:Berry1}
Berry, M.~V. and Klein, S. (1996).
\newblock Integer, fractional, and fractal talbot effects.
\newblock {\em Journal of Modern Optics}, 43(10):2139--2164.

\bibitem[Bluhm et~al., 1996]{rev:Bluhm1}
Bluhm, R., Kostelecky, V.~A., and Porter, J.~A. (1996).
\newblock The evolution and revival structure of quantum wave packets.
\newblock {\em American Journal of Physics}, 64:944.

\bibitem[Chen and Yeazell, 1998]{rev:Chen1}
Chen, X. and Yeazell, J.~A. (1998).
\newblock Analytical wave-packet design scheme: Control of dynamics and
  creation of exotic wave packets.
\newblock {\em Physical Review A}, 57(4):R2274--R2277.

\bibitem[Choi et~al., 2001]{bec:Choi1}
Choi, S., Burnett, K., Friesch, O.~M., Kneer, B., and Schleich, W. (2001).
\newblock Spatiotemporal interferometry for trapped atomic bose-einstein
  condensates.
\newblock {\em Physical Review A}, 63:065601.
\newblock cond-mat/0011468.

\bibitem[Cooper et~al., 2001]{misc:Cooper1}
Cooper, F., Khare, A., and Sukhatme, U. (2001).
\newblock {\em Supersymmetry in Quantum Mechanics}.
\newblock World Scientific Publishing Co.

\bibitem[Dubra and Ferrari, 1999]{opt:Dubra1}
Dubra, A. and Ferrari, J.~A. (1999).
\newblock Diffracted field by an arbitrary aperture.
\newblock {\em American Journal of Physics}, 67(1):87--92.

\bibitem[Friesch et~al., 2000]{car:Friesch1}
Friesch, O., Marzoli, I., and Schleich, W.~P. (2000).
\newblock Quantum carpets woven by wigner functions.
\newblock {\em New Journal of Physics}, 2(4):4.1--4.11.

\bibitem[Griffiths, 1995]{misc:Griffiths1}
Griffiths, D.~J. (1995).
\newblock {\em Introduction to quantum mechanics}.
\newblock Prentice Hall, Englewood Cliffs, N.J.

\bibitem[Grossmann et~al., 1997]{car:Grossman1}
Grossmann, F., Rost, J.-M., and Schleich, W.~P. (1997).
\newblock Spacetime structures in simple quantum systems.
\newblock {\em Journal of Physics A}, 30:L277--L283.

\bibitem[Hall et~al., 1999]{car:Hall1}
Hall, M.~J., Reineker, M.~S., and Schleich, W.~P. (1999).
\newblock Unravelling quantum carpets: A travelling wave approach.
\newblock {\em Journal of Physics A: Mathematics and General}, 32:8275--8291.
\newblock quant-ph/9906107 v2.

\bibitem[Jie et~al., 1998]{rev:Jie1}
Jie, Q.-L., Wang, S.-J., and Wei, L.-F. (1998).
\newblock Partial revivals of wave packets: An action-angle phase-space
  description.
\newblock {\em Physical Review A}, 57(5):3262--3267.

\bibitem[Kaplan et~al., 2000]{trace:Kaplan2}
Kaplan, A., Marzoli, I., Lamb, w.~J., and Schleich, W. (2000).
\newblock Multimode interference: Highly regular pattern formation in quantum
  wave-packet evolution.
\newblock {\em Physical Review A}, 61:032101.

\bibitem[Kaplan et~al., 1998]{trace:Kaplan1}
Kaplan, A., Stifter, P., van Leeuwen, W., Lamb, W.~J., and Schleich, W.~P.
  (1998).
\newblock Intermode traces-fundamental interference phenomenon in quantum and
  wave physics.
\newblock {\em Physica Scripta}, T76:93--97.

\bibitem[Knospe and Schmidt, 1996]{rev:Knopse1}
Knospe, O. and Schmidt, R. (1996).
\newblock Revivals of wave packets: General theory and application of rydberg
  clusters.
\newblock {\em Physical Review A}, 54(2):1154--1160.

\bibitem[Lock and Andrews, 1992]{opt:Lock1}
Lock, J.~A. and Andrews, J.~H. (1992).
\newblock Optical caustics in natural phenomena.
\newblock {\em American Journal of Physics}, 60(5):397--407.

\bibitem[Loinaz and Newman, 1999]{rev:Loinaz1}
Loinaz, W. and Newman, T. (1999).
\newblock Quantum revivals and carpets in some exactly solvable systems.
\newblock {\em Journal of Physics A: Mathematics and General}, 32:8889--8895.
\newblock quant-ph/9902039.

\bibitem[Marzoli et~al., 1998a]{car:Marzoli2}
Marzoli, I., Bialynicki-Biruli, I., Friesch, O., Kaplan, A., and Schleich, W.
  (1998a).
\newblock The particle in the box: Intermode traces in the propagator.
\newblock quant-ph/9804015.

\bibitem[Marzoli et~al., 1998b]{car:Marzoli1}
Marzoli, I., Saif, F., Bialynicki-Birula, I., Friesch, O., Kaplan, A., and
  Schleich, W. (1998b).
\newblock Quantum carpets made simple.
\newblock {\em Acta Physica Slovaca}, 48(3):323--333.
\newblock quant-ph/9806033.

\bibitem[Nieto and Simmons, 1979]{misc:Nieto1}
Nieto, M.~M. and Simmons, L.~J. (1979).
\newblock Limiting spectra from confining potentials.
\newblock {\em American Journal of Physics}, 47(7):634--635.

\bibitem[O'Connell, 2002]{O'Connell2002}
O'Connell, R. (2002).
\newblock A foray into quantum dynamics.
\newblock Amherst College Honors Thesis, quant-ph/0212092.

\bibitem[Razi~Naqvi et~al., 2001]{rev:Naqvi1}
Razi~Naqvi, K., Waldenstrom, S., and Haji~Hassan, T. (2001).
\newblock Fractional revival of wave packets in an infinite square well: a
  fourier perspective.
\newblock {\em European Journal of Physics}, 22:395--402.

\bibitem[Rozmej and Arvieu, 1998]{rev:Rozmej1}
Rozmej, P. and Arvieu, R. (1998).
\newblock Clones and other interference effects in the evolution of
  angular-momentum coherent states.
\newblock {\em Physical Review A}, 58(6):4314--4329.
\newblock quant-ph/9801018.

\bibitem[Ruostekoski et~al., 2001]{bec:Schleich1}
Ruostekoski, J., Kneer, B., Schleich, W.~P., and Rempe, G. (2001).
\newblock Inteference of a bose-einstein condensate in a hard-wall trap: From
  the nonlinear talbot effect to the formation of vorticity.
\newblock {\em Physical Review A}, 63:043613.
\newblock cond-mat/9908095.

\bibitem[Wright et~al., 1997]{bec:Wright1}
Wright, E.~M., Wong, T., Collett, M.~J., Tan, S.~M., and Walls, D.~F. (1997).
\newblock Collapses and revivals in the interference between two bose-einstein
  condensates formed in small atomic samples.
\newblock {\em Physical Review A}, 56(1):591--602.
\newblock cond-mat/9611211.

\end{thebibliography}

\end{document}